\documentclass[conference]{IEEEtran}
\IEEEoverridecommandlockouts
\usepackage{cite}
\usepackage{float}
\usepackage{array}
\usepackage{tabularx}
\usepackage{amsmath,amssymb,amsfonts}
\usepackage{algorithmic}
\usepackage{graphicx}
\usepackage{textcomp}
\usepackage{circledsteps}
\usepackage{xcolor}
\usepackage[caption=false,font=footnotesize]{subfig}
\usepackage{booktabs}
\usepackage{lmodern}         
\usepackage{url}             
\usepackage{enumitem}        
\usepackage{hyperref}        
\usepackage{subfig}

\hypersetup{
  colorlinks=true,
  linkcolor=black,
  citecolor=black,
  urlcolor=blue
}

\makeatletter
\def\bstctlcite#1{\@bsphack
  \@for\@citeb:=#1\do{\edef\@citeb{\expandafter\@firstofone\@citeb}%
    \if@filesw\immediate\write\@auxout{\string\citation{\@citeb}}\fi}%
  \@esphack}
\makeatother
\def\BibTeX{{\rm B\kern-.05em{\sc i\kern-.025em b}\kern-.08em
    T\kern-.1667em\lower.7ex\hbox{E}\kern-.125emX}}

\begin{document}
\bstctlcite{IEEEexample:BSTcontrol}
\title{AIIM: Adaptive Inter-cell Interference Mitigation for Heterogeneous Multi-vendor 5G O-RAN Networks \thanks{This material is based upon work supported by the National Science Foundation under Grant Numbers CNS-2202972, CNS- 2318726, and CNS-2232048.}
}

\author{\IEEEauthorblockN{Samuel Reinders, Alireza Ebrahimi Dorcheh, Ryan Barker, Tolunay Seyfi, Fatemeh Afghah}
\IEEEauthorblockA{\textit{Department of Electrical and Computer Engineering, Clemson University, Clemson SC USA} \\
\{sreinde,alireze,rcbarke,tseyfi,fafghah\}@clemson.edu}
}

\maketitle

\begin{abstract}
Inter-cell interference is a persistent issue in dense 5G deployments, especially in heterogeneous Open Radio Access Network (O-RAN) environments where coordination between base stations is limited. This paper presents AIIM, an adaptive inter-cell interference mitigation xApp for the O-RAN near-real-time RAN Intelligent Controller (near-RT RIC) that performs coordinated physical resource block (PRB) allocation across multiple base stations under diverse traffic demands and channel conditions.

Unlike prior studies that rely primarily on simulation or fully hardware-centric testbeds, AIIM is developed and evaluated in a full-stack O-RAN system built on srsRAN, Open5GS, and O-RAN Software Community (ORAN-SC), and deployed on a hybrid experimental platform that simultaneously combines software defined radio (SDR)-based and virtual gNodeBs (gNBs) and user equipment (UEs). This design preserves realistic PHY-layer interactions while substantially improving scalability, reproducibility, and cost-effectiveness for multi-cell interference experiments. AIIM explicitly models overlapping PRB regions across neighboring cells and learns coordinated allocation policies that adapt to per-user QoS demand and pathloss variation across the network. Experimental results show that AIIM improves QoS satisfaction and reduces interference-induced PRB loss relative to proportional-fair scheduling baselines while maintaining comparable aggregate network throughput. These results demonstrate the promise of scalable, learning-driven O-RAN control for practical interference management in heterogeneous multi-gNB 5G networks.\footnote{A video demonstration of the running system can be found at https://github.com/sireinders/AIIM-Multi-gNB-Interference.git.}
\end{abstract}


\section{Introduction} \label{Introduction}
Fifth-generation (5G) cellular networks support various services with diverse QoS requirements, while dense deployments are increasingly used to meet rising traffic demands\cite{interference_study}. In such environments, users experience highly variable channel conditions and interference depending on their location relative to neighboring base stations. These effects become more pronounced in complex heterogeneous deployments like multi-vendor neutral host networks (NHNs) and private 5G networks that commonly use shared or unlicensed spectrum, such as the Citizens Broadband Radio Service (CBRS) band~\cite{NHNs,CBRS_priv}. In these settings, competition or privacy concerns restrict inter-cell coordination and scheduling, increasing resource contention and inter-cell interference. For instance, without coordination, neighboring base stations may operate on the same carrier frequencies and independently schedule the same physical resource block (PRBs) during the same transmission interval\cite{interference_study}. Although Orthogonal Frequency-Division Multiple Access (OFDMA) preserves orthogonality within a cell, it does not prevent co-channel interference (CCI) across cells. This is particularly harmful for users near cell edges, where competing signals from adjacent base stations (gNBs) can significantly reduce throughput, increase block error rate (BLER), and violate quality of service (QoS) requirements. These challenges motivate interference mitigation mechanisms that improve spectrum utilization while enabling fair coordination across complex deployments\cite{interference_study}.

The Open Radio Access Network (O-RAN) framework provides a promising foundation for addressing these challenges in multi-vendor and heterogeneous deployments. O-RAN disaggregates traditional RAN functions and implements them in software, enabling more flexible network control under changing wireless conditions. By defining open and standardized interfaces, O-RAN enables the integration of intelligent control mechanisms through RAN Intelligent Controllers (RICs), allowing artificial intelligence (AI) and machine learning (ML) workflows to be embedded directly into the network control loop~\cite{ORAN_background}. The near-Real Time RIC (near-RT RIC) plays a central role in this architecture, providing control and optimization capabilities on timescales ranging from 10~ms to 1~s. Through the deployment of third-party xApps managed by the near-RT RIC, fine-grained and adaptive control of RAN behavior across base stations from different vendors is possible~\cite{ORAN_capabilities}. These capabilities make the near-RT RIC a suitable platform for implementing learning-based interference mitigation strategies.

Reinforcement learning (RL) has emerged as an effective paradigm for learning complex control policies in 5G and beyond networks, including slice-aware resource allocation, dynamic spectrum sharing, and RAN-level optimization~\cite{REAL, Spectrum_sharing, lotfi_DRL}. In RL-based approaches, agents observe key performance metrics (KPMs) from the wireless environment and receive feedback based on their actions in the form of rewards, enabling them to iteratively learn policies that optimize long-term performance objectives. Metrics such as aggregate throughput, spectral efficiency, transmission queue length, and block error rate (BLER) provide valuable insight into overall network behavior and serve as informative reward signals for learning-based interference mitigation.

Existing work on interference mitigation often emphasizes either simulation-scale evaluation or hardware realism, but rarely both within a deployable O-RAN control loop. This paper presents, a near-RT RIC xApp for coordinated cross-cell PRB allocation in heterogeneous 5G O-RAN networks, and validates it on a hybrid platform that combines software defined radio (SDR)-based and virtual gNBs and user equipment (UEs). By explicitly modeling shared PRB regions and adapting allocation decisions to user demand and pathloss, the proposed system enables scalable yet realistic experimental evaluation of inter-cell interference mitigation in multi-gNB deployments. The main contributions of this paper are as follows:
\begin{itemize}[leftmargin=*]
    \item We design AIIM, a near-RT RIC xApp for coordinated inter-cell interference mitigation through cross-cell PRB allocation in heterogeneous 5G O-RAN networks.
    \item We develop a hybrid experimental platform that combines SDR-based and virtual gNBs and UEs, enabling a practical balance between PHY-layer realism, scalability, and reproducibility.
    \item We formulate inter-cell interference mitigation as a learning-based control problem over shared PRB regions using user demand and pathloss as state information.
    \item We implement the full-stack system using srsRAN, Open5GS, and ORAN-SC, and experimentally compare AIIM against proportional-fair and learning-based baselines.
\end{itemize}

\section{Related Work}
Interference management has been a core research topic throughout the evolution of cellular networks. Higher user numbers have necessitated spatial and spectral re-use resulting in dense network deployments that, in certain cases, sacrifice network performance when neighboring cells unknowingly contend for resources. For instance, CCI occurs when adjacent cells reuse the same frequency resources simultaneously~\cite{Interference_review}. A common approach to this challenge focuses on equitable inter-cell PRB management. By defining a high-level policy that is aware of the network structure, PRBs can be allocated to cells in a fixed manner through frequency planning, or dynamically, using communication over the network backhaul such as enhanced inter-cell interference coordination~\cite{eICIC} and coordinated multi-point (CoMP) transmission~\cite{CoMP}. Although these techniques can reduce interference, they either sacrifice robustness or require strict synchronization and substantial backhaul signaling, making them less ideal for real-time PRB control~\cite{interference_study, co-channel}.

As an alternative, RL-based resource allocation algorithms have become popular for real-time, dynamic optimization of 5G O-RAN networks. REAL ~\cite{REAL} and DORA ~\cite{dora} demonstrate how RL-enabled xApps can be deployed on an open-source near-RT RIC to perform closed-loop slice-level PRB scheduling for meeting diverse user QoS requirements. Similarly, ORANSlice~\cite{oranslice} provides an open-source platform for slice-level resource management within O-RAN architectures.

Several recent works apply these learning-based approaches to interference mitigation. The ChARM framework~\cite{Spectrum_sharing} demonstrates the feasibility of real-time, data-driven cross-technology spectrum sharing with a deep neural network (DNN) model that leverages O-RAN control loops to reconfigure network parameters and avoid interference. The xDiff model~\cite{xDiff} introduces a diffusion-based RL approach for collaborative inter-cell interference mitigation (ICIM) by manipulating the per-user proportional fair (PF) metrics used by the resource schedulers of each cell. Evaluated on an over-the-air (OTA) testbed consisting of a single centralized unit (CU), three SDR distributed units (DUs), and ten commercial smartphones, xDiff outperforms state-of-the-art ICIM techniques by minimizing delay for users while achieving comparable throughput and BLER. Conversely, Qualcomm's Interference-Aware Intelligent Scheduling (IAIS) framework~\cite{qualcomm} uses an offline-trained Long Short-Term Memory (LSTM) model to drive intelligent link adaptation and modulation and coding scheme (MCS) selection to minimize interference within an OTA industrial Internet-of-Things (IIoT) factory automation testbed. MLCIMO~\cite{MLCIMO} proposes an ML-based classification model for nested macro/micro-cell deployments that characterizes interference levels and drives user-offloading to maximize network performance in a fully simulated environment.

Existing interference research and 5G O-RAN testbeds each have a critical limitation. First, to the best of our knowledge, there are no learning-based ICIM algorithms investigating multi-vendor 5G O-RAN deployments with different user service profiles like Enhanced Mobile Broadband (eMBB) and Massive Machine Type Communications (mMTC). Secondly, existing 5G O-RAN testbeds typically assume a single device-type in their network architectures. Namely, industry-leading large-scale wireless experimental platforms such as Colosseum~\cite{bonati2021colosseum}, AERPAW~\cite{panicker2021aerpaw}, and OAIC~\cite{msuoaic2023} are limited to software-only \emph{or} purely SDR-based architectures at one time. This limits the scalability of realistic wireless experiments to the size of the hardware testbed. Further, access to these platforms are limited and hardware resources are shared, making it difficult to conduct comprehensive experimentation in a timely manner. Fully simulated environments have been proposed to alleviate accessibility and scaling concerns; however, they neglect or omit important PHY-layer effects and protocol timing constraints that are required for properly emulating complex cellular deployments.

Mixed hardware/software device architectures offer a practical compromise that allow realistic, low-cost, reproducible experimentation within standard laboratory environments while scaling user density and traffic diversity. In this work, we propose an adaptive interference mitigation algorithm implemented within a multi-vendor O-RAN environment consisting of simultaneous SDR-based and virtual base stations and user devices with diverse QoS profiles. By combining real RF hardware with scalable virtual users, the proposed architecture enables cost-effective experimentation while capturing realistic PHY-layer hardware behavior. This design allows RL-based resource allocation and ICIM algorithms to be evaluated under diverse user demand and channel scenarios.

\section{System Model}
\subsection{Interference Modeling}\label{subsec:interference_model}
\begin{figure}
    \centering
    \includegraphics[width=0.8\linewidth]{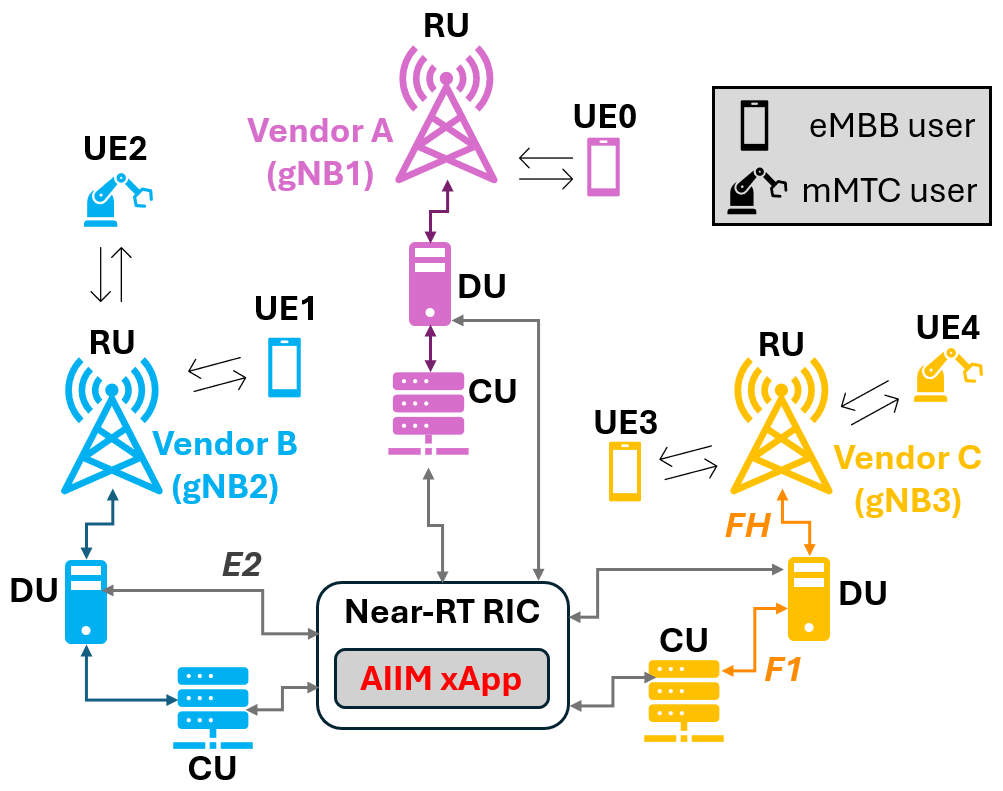}
    \caption{Virtual multi-vendor system model for per-cell PRB allocation including user slices and RU relative locations.}
    \label{fig:gnb_layout}
\end{figure}

In this section, we describe the system model and the assumptions adopted in this paper for interference analysis. We consider an O-RAN network composed of three vendor's base stations, each with their own radio unit (RU), distributed unit (DU), and centralized unit (CU). The DUs and CUs connect to a shared NRT-RIC via the E2 interface for metric reporting and network control. The open fronthaul interface (FH) and open midhaul interface (F1) connect the RU to DU and DU to CU, respectively. The virtual environment layout is illustrated in Fig. \ref{fig:gnb_layout}. Following the New-Radio (NR) standard, we define 52 PRBs per gNB, 10 MHz channel bandwidth, and 15 kHz subcarrier spacing in FR1. Vendor A's gNB supports an eMBB slice, while Vendors B and C support two slices, eMBB and mMTC.

\begin{figure}
    \centering
    \includegraphics[width=1\linewidth]{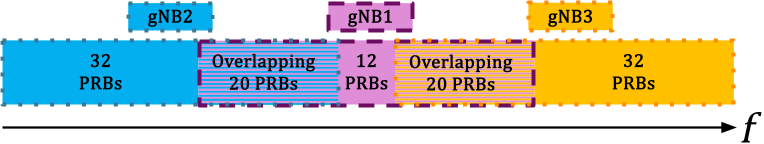}
    \caption{Frequency-domain PRB distribution for gNB2 (blue, virtual), gNB1 (pink, SDR-based), and gNB3 (orange, virtual).}
    \label{fig:prb_diagram}
\end{figure}

For analytical tractability, we consider inter-cell interference between users connected to different base stations based on overlapping frequency PRB assignments. If two separate base stations allocate the same frequency PRB to different users, the entire downlink bandwidth corresponding to that PRB is lost for both users. In our experiments, we do not consider partial-PRB overlap. The frequency-domain PRB overlap between the gNBs is defined in Fig. \ref{fig:prb_diagram}. There are a total of 156 PRBs, of which, there are 116 unique frequency PRBs.
The impact of interference is realized within the system through the following process. First, the algorithm determines per-user PRB allocations based on their respective allocation strategies defined in Section~\ref{sec:evaluation}. We assume that each gNB allocates their private (non-overlapping) PRBs to users before their shared (overlapping) PRBs. Therefore, the number of shared PRBs allocated by a single gNB is equal to the total PRBs it allocated subtracted by its private PRBs. The PRBs lost to interference ($\text{PRB}_L$) are calculated using  equation (\ref{eq:overlap}) for each shared PRB region the gNB has
\begin{equation} \label{eq:overlap}
\text{PRB}_L = {\max(0, (\text{PRB}_{S1} + \text{PRB}_{SX})-\text{PRB}_O})
\end{equation}
where $\text{PRB}_{S1}$ and $\text{PRB}_{SX}$ represent the number of shared PRBs allocated by gNB1 and its neighbor (gNB2 or gNB3), respectively, and $\text{PRB}_O$ denotes the total available PRBs in that specific overlapping frequency region. This value is then subtracted from the original PRB allocations to reflect effective throughput degradation for the users.

\subsection{Interference-Aware PRB Allocation RL Agent}\label{subsec:rl_form}
The inter-cell interference mitigation (ICIM) xApp employs an offline-trained proximal policy optimization (PPO) agent comprised of separate actor and critic networks, each having two fully-connected layers of 64 neurons. PPO uses on-policy learning to sample actions and actively refine its policy. PPO was selected due to its stable policy updates and strong performance in complex, dynamic environments. Finalized hyperparameters for the PPO model were determined through a parameter sweep of learning\_rate, gamma, ent\_coeff, n\_steps, batch\_size, and n\_epochs, in which we selected the parameters that achieved the strongest performance on validation scenarios without overfitting. The PPO and DQN model parameters are summarized in Table \ref{tab:ppo_param}.
\begin{table}[h!]
\centering
\caption{Model Parameters}
\label{tab:ppo_param}
\begin{tabular}{p{3.25cm} >{\centering\arraybackslash}p{1cm} >{\centering\arraybackslash}p{0.75cm} >{\centering\arraybackslash}p{0.75cm}}
\hline
\textbf{Parameter \footnotemark} & \textbf{Value} & \textbf{PPO} & \textbf{DQN} \\
\hline
\noalign{\vspace{2pt}}
learning\_rate & 3e-4 & $\checkmark$ & $\checkmark$ \\
gamma & 0.99 & $\checkmark$ & $\checkmark$ \\
ent\_coef & 0.01 & $\checkmark$ & $\chi$ \\
n\_steps & 512 & $\checkmark$ & $\chi$ \\
batch\_size & 16 & $\checkmark$ & $\checkmark$ \\
n\_epochs & 3 & $\checkmark$ & $\chi$ \\
exploration\_fraction & 0.1 & $\chi$ & $\checkmark$ \\
exploration\_initial\_eps & 1.0 & $\chi$ & $\checkmark$ \\
exploration\_final\_eps & 0.05 & $\chi$ & $\checkmark$ \\
\noalign{\vspace{1pt}}
\hline
\end{tabular}
\end{table}
\footnotetext{Remaining parameters set to StableBaseline3 default values.}

The RL state space $\mathcal{S}$ consists of concatenated user state vectors $s_t = (c_{user}, \lambda_t, L_t)$ at time step $t$, where the components are defined as:
\begin{itemize}
    \item $c_{user} \in \mathcal{C}$: QoS category, where $\mathcal{C} = \{c_\text{eMBB}, c_{\text{mMTC}}\}$.
    \item $\lambda_t \in \mathbb{R}_{\geq 0}$: Throughput demand in Bytes per second (Bps).
    \item $L_t \in \mathbb{R}$: Channel pathloss in decibels (dB).
\end{itemize}
The complete state space is thus defined by the Cartesian product $\mathcal{S} = \mathcal{C} \times \mathbb{R}_{\geq 0} \times \mathbb{R}$. During preliminary testing, this lightweight state design enabled faster policy convergence and  better reward performance than those that included historic demand or resource allocations of users. We hypothesize that the agent struggles to learn high-level network trends when overwhelmed with dense, highly-dependent per-user metrics.

The action space $\mathcal{A}$ defines a discrete PRB splits for each shared PRB region. This is done to enable dynamic and collaborative bandwidth sharing between base stations without redundant backhaul signaling. Within a single shared region, a lower-priority gNB selects the number of PRBs it will defer to usage by a higher priority gNB $a_t \in \mathcal{A}$, where $\mathcal{A} = \{0, 1, 2, ..., K\}$. $K$ is the maximum number of overlapping PRBs. By selecting an action, the RL agent determines each cell's share of the total PRBs. After this, individual base stations distribute their PRBs proportionally between users based on their demand. In our system, gNB1 is configured with the highest priority. This approach prevents action space explosion inherent to fine-grained per-user PRB assignments as user counts scale, while maintaining flexibility to network conditions.

The reward function is designed in such a way for the agent to generate an effective policy that dynamically adjusts resource allocations across multiple cells as to \Circled{1} satisfy per-user QoS demands and \Circled{1} achieve efficient spectrum utilization amidst inter-cell interference conditions. Within the environment, let $\mathcal{N}$ denote the set of all cells and $\mathcal{U}$ represent the set of all UEs. An overlapping PRB region between two cells A and B is denoted as $\mathcal{O}_{A,B}$ with the set of all overlapping regions written as $\mathcal{O}$.

For UE $i \in \mathcal{U}$, we use the clipped ratio of average achieved throughput $\lambda_A$ to requested throughput $\lambda_R$ in Bytes to represent the per-user satisfaction reward as $r_s(t)$:
\begin{equation} \label{eq:user_reward}
r_s(t) = \min({\frac{\lambda_A}{(\lambda_R + \epsilon)}}, 1)
\end{equation}
where $\epsilon = 1 \times 10^{-6}$. This ratio-based throughput reward design consistently outperformed preliminary methods that used a sigmoid function or PRB-based rewards which caused the agent to settle into sub-optimal policies due to saturating rewards or coarse reward signals. Our approach discourages the agent from resource over-allocation, while accounting for the wireless channel and providing a smooth gradient for learning.

During our initial testing, we found that relying exclusively on user-centric rewards induced myopic greedy policies. Instead of a globally equitable strategy, cells with fewer users or higher interference were starved of shared resources entirely. To combat this, we incorporated a spectrum utilization reward $r_u(t)$ for each $\mathcal{O}_{A,B}$ that can be written as:
\begin{equation} \label{eq:gnb_reward}
r_u(t) = {\frac{n*\phi*\alpha}{\chi}}
\end{equation}
where $n$ is the total number of users within the network, $\phi$ is a configurable fairness parameter, $\chi$ is the size of $\mathcal{O}_{A,B}$ in PRBs, and $\alpha$ is the number of shared PRBs allocated to the higher priority gNB. We select $\phi$ to be 0.05 in our experiments, after values of 0.1, 0.025 and 0.01 produced 13\%, 3\%, and 7\% higher QoS violations in evaluation scenarios. With this utilization reward, the agent learned to better share resources between vendor's cells and avoid selfish behaviors while maintaining overall network performance in highly dynamic user scenarios.

The total reward for the agent across all users and overlapping PRB regions is shown as:
\begin{equation} \label{eq:global_reward}
r(t) = {\sum_{i \in \mathcal{U}}{r_s(t)} + \sum_{j \in \mathcal{O}}{r_u(t)}}
\end{equation}

\subsection{Training Environment and Traffic Model}\label{subsec:training}
The reinforcement learning agent was trained offline in a Gymnasium-based environment designed to emulate the multi-cell interference scenarios observed in the testbed. Training was performed over 10,000 episodes with 100 steps per episode. At the beginning of each episode, user positions and bearing are randomly initialized within the coverage area of their serving base station. User mobility follows the 3GPP Extended Vehicular A (EVA) channel model with a velocity of 10 m/s~\cite{3gpp_tr38913}. User demand profiles are randomly sampled from ranges corresponding to realistic eMBB and mMTC service classes for every step. Table~\ref{tab:service_profiles} summarizes the traffic profiles used during training. Demand ranges were selected based on achievable downlink throughput observed on the experimental platform while producing a comprehensive range of interference conditions (none - high) and maintaining UE connectivity.
\begin{table}[h!]
\centering
\caption{Service Profiles}
 \label{tab:service_profiles}
\begin{tabular}{p{2.68cm} >{\centering\arraybackslash}p{2.1cm} >{\centering\arraybackslash}p{1.3cm} >{\centering\arraybackslash}p{1.15cm}}
\hline
\noalign{\vspace{1pt}}
\textbf{Parameter} & \textbf{Demand Range} & \textbf{Frequency (1/s)} & \textbf{PRB Range} \\
\hline
\noalign{\vspace{3pt}}
SDR eMBB (UE0) & 1.5MB - 3.5MB & 1 & 20 - 49 \\
VIR eMBB (UE1,3) & 907kB - 3.1MB & 1 & 12 - 44 \\
VIR mMTC (UE2,4) & 37.5kB - 62.5kB & 4 & 5 \\
\hline
\end{tabular}
\end{table}
Minimum connectivity requirements were determined empirically during preliminary testing. A minimum allocation of five PRBs was required to maintain stable UE connectivity, corresponding to a minimum PRB allocation ratio of 10\% within the RIC resource control policy.

\section{Experimental Evaluation}\label{sec:evaluation}
\subsection{Testbed Architecture}
\begin{figure}[H]
    \centering
    \includegraphics[width=1.0\linewidth]{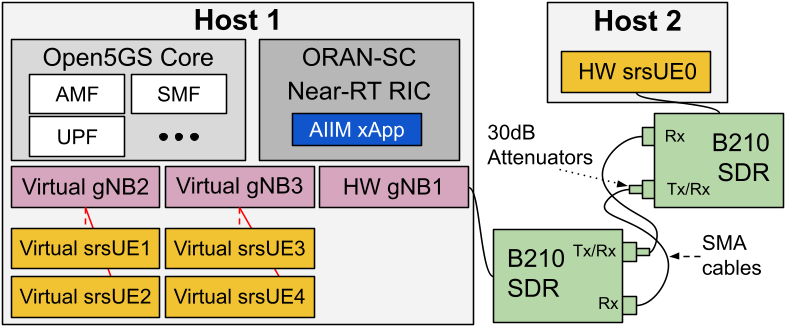}
    \caption{Testbed architecture with hardware connections and software processes.}
    \label{fig:mh_setup}
\end{figure}

The experimental platform integrates the srsRAN 5G stack, the Open5GS core network, and the O-RAN Software Community (OSC) RIC to create a hybrid multi-gNB and multi-UE environment for evaluating PRB resource allocation under multi-cell co-channel interference. The architecture combines SDR-based and virtual network elements to enable scalable experimentation while maintaining realistic PHY-layer interactions. The system consists of the following components:
\begin{itemize}[leftmargin=*]
    \item \textbf{Open5GS Core}: Provides 5G core network functions for UE registration (AMF), session management (SMF), and user-plane forwarding (UPF).
    \item \textbf{ORAN-SC Near-RT RIC}: Hosts the ICIM xApp responsible for coordinating PRB allocation decisions across multiple gNBs via the E2 interface.
    \item \textbf{srsRAN gNB}: Implements a monolithic gNB architecture with RU, CU, and DU functionality. Both SDR-based and virtual ZeroMQ (ZMQ)-based gNB instances are deployed. 
    \item \textbf{srsRAN UE}: Provides both SDR-based and virtual user equipment implementations supporting LTE and 5G Non-Stand Alone (NSA) connectivity.
    \item \textbf{GNU Radio Blocks}: Used to multiplex and route I/Q streams between virtual users and gNB instances using ZMQ sockets. 
\end{itemize}
The testbed was composed of two physical hosts (Host 1 and Host 2), each connected to an Ettus B210 SDR. Host 1 executed the Open5GS core network and ORAN-SC Near-RT RIC within Docker containers, while the SDR-based base station (gNB1), two virtual base stations (gNB2 and gNB3), and four virtual srsUEs instances (UE1-UE4) were executed as processes on the same host. Host 2 executed a single SDR-based UE process (UE0) connected to its local B210 SDR. GNU Radio flowgraphs were used to route and multiplex I/Q samples for the virtual UEs, enabling successful registration with the Access and AMF, and Protocol Data Unit (PDU) session establishment through the Session Management Function (SMF). The trained ICIM agent was deployed as an xApp running on the ORAN-SC Near-RT RIC integrated with the srsRAN stack. Downlink traffic was generated using iPerf3 and performance metrics were reported to the Near-RT RIC via the E2 interface. A high-level overview of the testbed architecture is shown in Fig. \ref{fig:mh_setup}.
\subsection{Experimental Configuration}
The experimental testbed operates in the n3 frequency band, which uses frequency division duplexing (FDD). The downlink and uplink center frequencies were configured at 1842.5 MHz and 1747.5 MHz, respectively. Due to limitations in GNU Radio-based multi-UE emulation with srsRAN, direct RF-level pathloss modeling is not feasible. Instead, large-scale fading effects were emulated using a calibrated throughput scaling model based on a reference pathloss value of $\text{PL}_\text{ref}=83.3 \text{dB}$. The calibrated throughput $\lambda_{cal}$ is defined as:
\begin{equation} \label{eq:throughput_scaling}
    \lambda_{cal} = \lambda_{\text{ach}} \cdot \max\left(g_{\min}, \min\left(1, e^{-\kappa(\text{PL}_{\text{UE}} - \text{PL}_{\text{ref,dB}})}\right)\right)
\end{equation}
$\lambda_{ach}$ represents the throughput achieved on the testbed before pathloss scaling, $g_\text{min} = 0.5$ ensures a minimum scaling factor to maintain UE connectivity, $\kappa = 5.5 \times 10^{-2}$ is a decay factor controlling signal attenuation, and $\text{PL}_\text{UE}$ represents the pathloss calculated from the UE position and downlink channel frequency.
\subsection{Evaluation Methodology}
To evaluate performance, we compared our model against four baselines using two performance metrics: 
\begin{itemize}
\item Total QoS violations across all users
\item Total PRBs lost due to inter-cell interference
\end{itemize}
QoS violations occur when a user's achieved throughput is less than it requested. These metrics jointly capture user-level performance and overall spectral efficiency of the network.

Evaluation was conducted over 5 episodes consisting of 70 environment steps each, for a total of 350 evaluation steps. While this sample size is constrained by the real-time execution requirements of the hardware-in-the-loop (HITL) testbed, the randomized initialization of UE positions and traffic demands in each episode ensures that the results capture a statistically diverse range of interference scenarios. During each step, downlink traffic was generated using iPerf3 for 15 seconds. Achieved throughput values were reported to the Near-RT RIC via the E2 interface and computed as the average of 10 throughput samples per user before pathloss scaling. User and demand scenarios remained consistent between evaluated models for a fair comparison. Each model was evaluated independently in continuous testing blocks.

\subsection{Baseline Algorithms}
The proposed PPO-based interference mitigation model was compared against four baseline approaches:
\begin{itemize}[leftmargin=*]
    \item \textbf{Non-Slice Aware PF (NSA-PF)}: Each DU independently allocates PRBs without inter-cell coordination.
    \item \textbf{Slice Aware Constant-Allocation PF (SA-CA-PF)}: Non-overlapping PRBs are statically partitioned among gNBs while slice-level demand is aggregated across the network. PRBs are evenly distributed among users belonging to the same slice.
    \item \textbf{Slice-Aware Variable-Allocation PF (SA-VA-PF)}: Non-overlapping PRBs are statically partitioned among gNBs with PRBs assigned according to the standard rate-to-average-rate ratio utility while respecting slice-level PRB guardrails $(N_s^{\min}, N_s^{\max})$.
    \item \textbf{Deep Q-Learning Network (DQN)}: An off-policy value-based reinforcement learning algorithm that replaces PPO and uses a 1-dimensional action space.
\end{itemize}
\subsection{Evaluation Results}
\begin{figure*}[!t]
    \centering
    \subfloat[Total PRBs lost to inter-cell interference]{
        \includegraphics[width=0.3\textwidth]{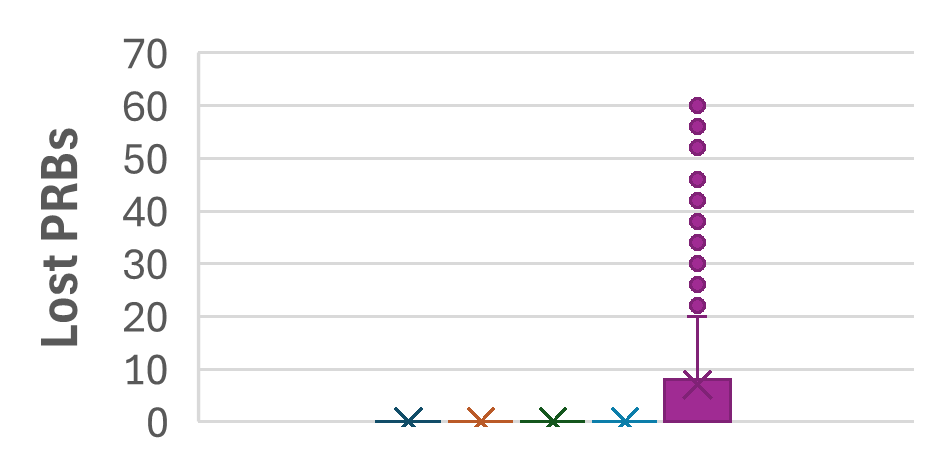}
        \label{fig:lost_prbs}
    }
    \hfil
    \subfloat[Total user QoS violations]{
        \includegraphics[width=0.3\textwidth]{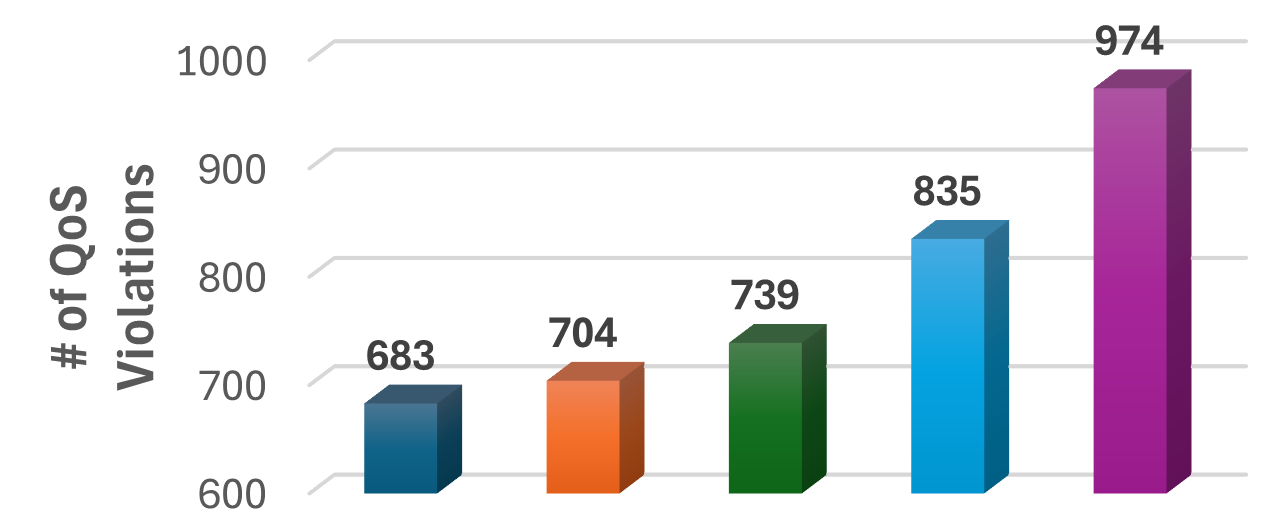}
        \label{fig:qos}
    }
    \caption{Performance metrics across PPO, DQN, SA-VA-PF, SA-CA-PF, and NSA-PF.}
    \label{fig:combined_performance}
\end{figure*}

The resulting PRBs lost to interference and user QoS violations across all the evaluation scenarios are displayed in Figs. \ref{fig:combined_performance}\subref{fig:lost_prbs} and \ref{fig:combined_performance}\subref{fig:qos}.
As expected, the NSA-PF algorithm is the only baseline that experiences PRB losses due to inter-cell interference. This happens because NSA-PF allows overlapping PRB usage across neighboring cells without coordination. In contrast, the other approaches prevent interference by statically or dynamically partitioning overlapping spectrum regions.

The maximum \emph{effective} PRBs loss in the system is 80 PRBs, which occurs when all three gNBs simultaneously allocate 52 PRBs and maximize over-utilization of the shared spectrum regions. During evaluation, NSA-PF experienced an average number of 7.08 lost PRBs with a standard deviation of 11.32. This variation across steps is caused by the diverse combinations of traffic profiles and user positions, which produce different interference patterns. Since each PRB equates to approximately 550 kbps of achievable throughput when MCS is at its maximum, losing seven PRBs results in an average throughput degradation of approximately 3.85 Mbps or roughly 6\% of the maximum aggregate network throughput. This result clearly illustrates the performance impact of interference and the need for effective mitigation strategies in dynamic networks.

While SA-VA-PF and SA-CA-PF effectively eliminate interference by partitioning spectrum resources between neighboring cells, they remain inflexible to changing user demands and channel conditions. The PPO and DQN models provide more robust solutions to interference by interpreting path-loss and user demand to dynamically adjust resource allocation decisions. This leads to better QoS satisfaction in complex multi-cell environments. Specifically, the PPO and DQN models reduce total QoS violations by 7.6\% and 4.7\%, respectively, when compared to the best-performing non-learning baseline algorithm. 

Table~\ref{tab:agg_tpt} summarizes the average aggregate network throughput achieved by each model, which are similar across evaluation scenarios.
\begin{table}[h!]
\centering
\caption{Average Aggregate Network Throughput}
 \label{tab:agg_tpt}
\begin{tabular}{m{0.8cm} >{\centering\arraybackslash}m{1.2cm}>{\centering\arraybackslash}m{1.45cm}>{\centering\arraybackslash}m{1.5cm}>{\centering\arraybackslash}m{0.6cm}>{\centering\arraybackslash}m{0.6cm}}
\hline
\noalign{\vspace{1pt}}
\textbf{Model} & NSA-PF & SA-CA-PF & SA-VA-PF & DQN & PPO \\
\hline
\noalign{\vspace{1pt}}
\textbf{Tput [Mbps]} & 42.69 & 43.92 & 52.72 & 50.31 & 50.18 \\
\noalign{\vspace{1pt}}
\hline
\end{tabular}
\end{table}
PPO and DQN models achieve slightly lower aggregate throughput than SA-VA-PF because they allocate additional PRBs to users experiencing higher pathloss in order to satisfy QoS requirements. This behavior improves fairness across users by ensuring reliable service even for users located near cell edges or experiencing unfavorable channel conditions.

Overall, these results indicate that learning-based interference mitigation strategies outperform static or semi-static rule-based approaches in dynamic multi-cell environments. PPO achieves the strongest performance due to its stable policy updates and improved exploration capabilities in stochastic environments, enabled by clipped policy gradients and on-policy optimization.

\section{Conclusion}
In this work, we proposed AIIM, an offline-trained ICIM xApp for heterogeneous multi-vendor 5G O-RAN networks. AIIM leverages reinforcement learning to determine per-cell resource allocation actions that minimize user QoS violations and avoid losing PRBs to interference, while maintaining near-optimal aggregate network throughput. AIIM was implemented on a mixed 5G testbed composed of virtual and hardware base stations and users, confirming its applicability toward real-world deployments. This enables future work that can leverage the high scalability of such systems while balancing implementation costs. Experimental results indicate that learning-based algorithms like AIIM's PPO model as well as the DQN baseline model readily outperform traditional static and semi-static rule-based interference mitigation techniques across unique user placement and demand scenarios. These findings highlight the potential of the O-RAN framework to support scalable interference mitigation solutions for heterogeneous multi-vendor networks.

\bibliographystyle{IEEEtran}
\bibliography{ref.bib}

\end{document}